\documentclass[12pt]{article}
\usepackage{graphicx}
\title{A Newtonian pre-introduction to gravitational lenses}

\author{T. Garel \footnote{Member of CNRS}\\
Service de Physique Th\'eorique, CEA/DSM/SPhT\\
Unit\'e de recherche associ\'ee au CNRS\\
91191 Gif sur Yvette Cedex, France\\}

\def\be{\begin{equation}}
\def\bea{\begin{eqnarray}}
\def\ee{\end{equation}} 
\def\eea{\end{eqnarray}} 
\begin{document}
\maketitle
\vskip 4mm

{\bf Abstract:} Understanding the deflection of light by a massive
deflector, as well as the associated gravitational lens phenomena,
require the use of the theory of General Relativity. I consider here
a classical analogy, based on Newton's equation of
motion for massive particles. These particles are emitted by a
distant source and deflected by the gravitational field of a (opaque)
star or of a (transparent) galaxy. The dependence of
the deviation angle $D$ on the impact parameter $b$, and the
- Euclidean - geometry of the (source, deflector, earth) triplet, imply
that different particle trajectories may reach an earth based
observer. Since $D(b)$ does not depend on the mass of the
particles, a (Newtonian) flavor of gravitational lenses
phenomena is naively obtained by setting the particles' velocity equal
to the speed of light. Orders of magnitude are obtained through this
classical approach, and are compared to the General Relativity results.


\medskip

\vskip 4mm
\hfill\mbox{Saclay T03/171}\\ \noindent \mbox{ }\\

\section{Introduction}
\label{intro}
Gravitational lenses phenomena
\cite{Mis_Tho_Whe,Sch_Ehl,Zak_Saz,Bou_Kan,Adl_Bar,Wamb} rest on the 
gravitational deflection of light, and their explanation require the
use of General Relativity (the gravitational deflection of light by 
the sun, of order $10^{-5}$ rd, was indeed a major issue at the very
beginning of the theory). This theory uses a rather heavy technical
machinery, which is not easily accessible to undergraduate
students. One of the motivations of the present paper is to offer,
through a classical approach, a feeling (and orders of magnitude) for
gravitational lenses phenomena.

Apart from extreme cases (neutron stars, black holes,...), the
gravitational defection of light in
General Relativity (GR) is small, and can be described by a weak gravity
approximation \cite{Mis_Tho_Whe,Sch_Ehl,Zak_Saz}. For quasistationary
isolated mass distributions, the evolution of the unit tangent vector
$(\vec e(s))$ to a light ray ($\vec r(s)$) is given to lowest order
(see below) by
\be
\label{grav0}
{d\vec e(s) \over ds}=-{2 \over c^2}   \vec {\nabla}_{\perp}U
\ee
where $c$ is the speed of light, $U$ the gravitational potential
created by static mass distributions, and $\bigl(\vec {\nabla}_{\perp}U=\vec
{\nabla}U -\vec e(\vec e \cdot \vec {\nabla}U)\bigr)$. 
The weak gravity approximation corresponds to ${U \over c^2}<<1$, and 
``lowest order''  means that the typical velocity $v$ of the mass
distribution is small compared to $c$: A more rigorous calculation
indeed shows that there appear other terms in the r.h.s of equation
(\ref{grav0}), among which the lowest order term is of order ${\rm
O}({v \over c}))$ (see eq. 4.17, p. 124 of \cite{Sch_Ehl}). 

\vskip 2mm 

Equation (\ref{grav0}) bears some resemblance to a classical
equation of motion; it is indeed a familiar remark that Newton's
classical equation for a massive particle in the gravitational field
of a deflector leads to a mass independent deflection angle (identity
of inertial and gravitational masses). In this paper, we study the
classical mechanical problem defined by the triplet (particle source
(S), deflector ($\Delta$), earth based observer (E)). An important
ingredient of this study is the dependence of the gravitational
deflection angle $D$ on the impact parameter $b$ of the
particles. Since the (Euclidean) distances between (S), ($\Delta$) and
(E) are finite, the exact calculation of $D(b)$ is rather involved,
even if one takes advantage of the central character of the
gravitational force and of some invariant properties (we use here the
Runge-Lenz vector and tensor).  A rather quantitative approach is used
here to derive the main features of the deflection $D(b)$ for the case
of spherical deflectors (mass $M$, radius $R$). 

To make contact with the (GR) weak gravity approximation, we consider
a similar approximation for the classical mechanical study, namely we
consider the limit (Gravitational energy 
$<< $ Kinetic energy). Further, if ($v_{\infty}$) is a measure of the 
particle velocity as it leaves the source, one has at some point to
make the (delicate) correspondence $v_{\infty} \to c$ in the classical
mechanical problem. In this paper we follow this view of the
gravitational deflection of light to obtain {\it orders of magnitude}, 
using typical undergraduate skills. It is shown
that, in some cases, several particle trajectories may reach observer
(E). These trajectories translate into a
basic gravitational lensing effect: 
the source (S) may give multiple (photonic) signals for observer
(E). The time delay between the reception of these different
trajectories (or signals) is also considered.

The plan of the paper is as follows. The gravitational action of the 
deflector, such as the sun (opaque star) or the Milky Way (transparent
galaxy), is studied in Section \ref{orders}. For simplicity, we assume
there that the source (S) and the observer (E) are both at
infinity.   When (S) and (E) are at a finite distance from the
deflector, a little bit of geometry shows that that (S) may have
multiple ``images'' (Section \ref{geometry}). Finally, an estimate of
the time lag between the reception of these ``images'' is obtained,
via the analogy of particle mechanics and geometrical optics (Section
\ref{time}).

\section{Study of the classical mechanical deflection problem}
\label{orders}
We consider a source (S) which emits, in an isotropic fashion,
isovelocity particles ($\vert \vec v \vert =v_{\infty}$) of mass
$m$. Unless otherwise stated, we will consider a point source. We
focus our interest on the deflector and take the source and the
observer at infinity. Let $\vec v_{\infty}=v_{\infty} \vec u_i$ be the
initial velocity of a test particle. The deflector ($\Delta$) may be
either opaque or transparent to the particles. The former case will be
illustrated by the sun (radius $R_{\odot} \simeq 7 \cdot 10^8$m and
mass $M_{\odot} \simeq 2 \cdot 10^{30}$kg), and the latter by a
spherical version of the Milky Way \footnote{see e.g.

http://hypertextbook.com/facts/2000/AlinaVayntrub.shtml;

http://www.stdimension.de/int/Cartography/mwtour.htm} (radius $R_G \simeq 2 \cdot 10^4$ light years (ly) \footnote{One light year
represents a distance of approximately $9.5 \cdot 10^{15}$m.} and mass
$M_G \simeq 10^{12} M_{\odot}$).

\bigskip

\subsection{The opaque detector}
The impact parameter $b$ and deviation angle $D$ are defined 
in Figure 1. The study of the classical Kepler problem  for
$b>R$ \footnote{ We do not 
make the distinction between the impact parameter ($b$) and the minimum
distance approach ($r_0$). The exact relation $r_0=b\left({1-{\rm
sin}{D \over 2} \over {\rm 
cos}{D \over 2}}\right)$ shows that for small $D$, one has $r_0 \simeq 
b$.} can be found in standard textbooks (see e.g. \cite{LL}). If
$({\vec u_i})$ and $({\vec u_f})$ are the unit vectors along initial
and final velocities, we have $\cos D(b) ={\vec u_i} \cdot {\vec u_f}$.

The Runge-Lenz vector of the Kepler problem reads 
\be
\label{RungeLenz}
\vec A=\vec v \times \vec L- {\cal G}Mm {\vec r \over r}
\ee
where $\vec r$ is the particule position measured from the center of
the deflector ($\Delta$), $\vec L=\vec r \times m\vec
v$ its angular momentum, and $\cal G$ the universal gravity
constant. As is easily checked, $\vec A$ is an invariant vector of the 
motion and writing $\vec A_i=\vec A_f$ \cite{Basa_Bian} leads to 
\be
\label{devia1}
{\rm {tan}} {D \over 2}= {{\cal G}M \over bv_{\infty}^2}
\ee
The weak gravity condition means that the potential energy $m U(b)$ is
a small part of the total energy, that is ${{\cal G}M \over
bv_{\infty}^2}<<1$. The deviation is then given by
\be
\label{devia2}
D(b) \simeq {2{\cal G}M \over bv_{\infty}^2}
\ee

The (GR) result is obtained from equation (\ref{grav0}) as $D_{GR}(b)
\sim {4{\cal G}M \over bc^2}$. Applying these results to the sun at
grazing incidence yields $D_{GR}(R_{\odot}) \sim 8 \ 10^{-6}$ rd.

Two remarks are in order (i) The naive identification
($v_{\infty} \to c$), yields an expression which is half the exact (GR)
result (ii) In classical mechanics, energy conservation implies that
$v^2(r)=v^2_{\infty} +{2{\cal G}M \over r}$. If one identifies $v_{\infty}$
with $c$, the velocity $v(r)$ is greater than $c$, the (weak) correction term
being of order ${{\cal G}M \over rc^2}$. For small $D$, the minimum
distance $r_0$ approach is of order $b$, leading to a maximum
supraluminal correction for the velocity of order ${{\cal G}M \over
bc^2}$, that is small. As a temporary conclusion, I would
say that for practical purposes, the weak gravity limit of the
classical case, together with the identification $v_{\infty}\to c$,
yields a correct order of magnitude for the (GR) value. Some
inconsistencies of classical mechanics appear in this identification,
but are small in the weak gravity limit. 

\subsection{The transparent deflector}

A preliminary remark is that $D(b)$ is not a monotonous function,
since it vanishes both for $b=0$ and $b=\infty$.
For ($b>R$), the deviation is exactly given by equation
(\ref{devia1}), the weak gravity approximation corresponding to equation
(\ref{devia2}).

The case ($b <R$) requires the study of two distinct phases of the
motion, since the gravitational field $\vec g(\vec r)$
on the test particle has different expressions, depending on the
particle position $\vec r$. We respectively denote these phases by
(out) and (in). Their respective contribution to the total deviation
$D(b)$ will be denoted by $D_{(out)}$ and $D_{(in)}$. We further
assume that the deflecting galaxy is homogeneous, and neglect all
collisions with the particles in phase (in).

In phase (in) of the motion, Gauss' theorem gives 
\be
\label{newton1}
\ddot {\vec r}=\vec g_{(in)}(\vec r)=-{{\cal G}M \over R^3}\vec
r=-\omega_0^2 \vec r
\ee
This harmonic motion has period $T_0={2\pi \over
\omega_0}=2\pi \sqrt{R^3 \over {\cal G}M}$. The
trajectory inside the deflector is elliptic or 
partially elliptic. One should express the initial and final boundary
conditions to get the deflection $D_{(in)}(b)$. These boundary conditions
depend on phase (out) of the trajectory ($r>R$), where 
Newton's equation reads
\be
\label{newton2}
\ddot {\vec r}=\vec g_{(out)}(\vec r)=-{{\cal G}M \over r^3}\vec r
\ee
The total deviation $D(b)$ is given by 
\be
D(b)=D_{(in)}(b)+D_{(out)}(b)
\ee

\subsubsection{Phase (out): ($r>R$)}

\bigskip

{\bf {(a) Rigorous solution}}

\bigskip

We want to calculate the deviation between initial
(source S) and final (entry into deflector ($\Delta$)) points. The final
point M has position $\vec r_{in}=\vec {OM}$ and velocity $\vec
v_{in}$ (Figure 2(a)), with $\vert\vec r_{in}\vert=R$, and $\vec
v_{in}=v_0 \vec u_{in}$. Energy conservation yields 
$v_0^2=v_{\infty}^2+2{{\cal G}M \over R}$. The deviation
$D_{out}^{(1)}$ for this part of the (out) trajectory is  
given by $\cos D_{out}^{(1)}=\vec u_{i} \cdot \vec u_{in}$.

Equating the projections of the Runge-Lenz vector (eq. (\ref{RungeLenz}))
along $\vec u_i$ for the initial (S) and final (M ) points leads to
\be 
\label{RungeLenz2}
{\cal G}Mm= \vec u_i \cdot \bigl(\vec v_{in} \times \vec L-{\cal
G}Mm {\vec r_{in} \over R}\bigr)
\ee

Defining $\cos (\pi-\Phi)=\vec u_i \cdot {\vec r_{in} \over R}$
(Figure 2(a)), we get
\be 
\label{dev1}
\sin D_{out}^{(1)} ={{\cal G}M \over {bv_0v_{\infty}}}(1-\cos \Phi)
\ee

The exact solution of the Kepler problem (see eq (14,7), p.46 of
\cite{LL}) yields
\be
\label{phi2}
\Phi=-\arccos \bigl({{b \over R}-d \over {\sqrt{1+d^2}}}\bigr)+\arccos
\bigl({-d \over {\sqrt{1+d^2}}}\bigr)
\ee
with $d={{\cal G}M \over bv_{\infty}^2}$.

Taking into account the symmetrical ($\Delta E$) contribution
(deviation $D_{out}^{(2)}$) finally gives

\be
\label{dout}
D_{out}(b)=2\arcsin \bigl({{\cal G}M \over {bv_0v_{\infty}}}(1-\cos
\Phi)\bigr) 
\ee
where $\Phi$ is given by (\ref{phi2}).

\bigskip

{\bf {(b) Weak gravity approximation}}

\bigskip

As previously stated, we expect gravitational deviations to be
weak. Equation (\ref{dout}) shows that the weak gravity (``small
${\cal G}$'') limit can be obtained by setting $d=0$ in
eq. (\ref{phi2}). The weak gravity approximation therefore reads 
\be
\label{dout2}
D_{out}(b) \simeq 2 \ {{\cal G}M \over {bv_{\infty}^2}}(1-\sqrt{1- {b^2
\over R^2}})
\ee
In particular, the small $b$ limit is given by $D_{out}(b) \sim {b \over
f}$, with $f={R^2v_{\infty}^2 \over {\cal G}M}$.

\subsubsection{Phase (in): ($r<R$)}

\bigskip

{\bf {(a) Rigorous solution}}

\bigskip

The geometry of the (in) phase is shown in Figure 2 (b). The angle
$\alpha$ is the angle between the radius vector $\vec r_{in}$ and the
velocity $\vec v_{in}$ as the particle enters the deflector (Figure
2(a)). From the conservation of the angular momentum, one has
\be
\label{angular}
mv_{\infty}b=mv_0R \sin \alpha
\ee
where $v_0^2=v_{\infty}^2+2{{\cal G}M \over R}$.

To calculate the deviation $D_{in}(b)$ between the entry and exit
points, one may solve the harmonic motion of eq. (\ref{newton1}). 
A more convenient way is to use the Runge-Lenz matrix invariant
associated with the harmonic oscillator
\be
\label{RungeLenz3}
{\bf A}={m \over 2}(\omega_0^2 \bf r \bf r+\bf v \bf v)
\ee
Starting from the invariance of ${\bf A}$, simple
calculations \cite{Siva} show that 
\be
\label{din2}
D_{(in)}(b)=2(\alpha -\beta)
\ee
where ${\rm sin}\alpha ={bv_{\infty} \over Rv_0}$, and
${\rm tan} 2\beta={v_0^2{\rm sin} 2\alpha \over {\omega_0^2 R^2+v_0^2
{\rm  cos}2\alpha}}$.  

\bigskip

{\bf {(b) Weak gravity approximation}}

\bigskip

In this approximation, we obtain
$\sin \alpha \simeq {b \over R}$, and
\be
\label{din3}
D_{in}(b) \simeq \tan D_{in}(b)=\tan 2(\alpha -\beta) \simeq {{\cal G}M \over
Rv_{\infty}^2} \sin 2\alpha
\ee

\subsubsection{Conclusion on the transparent deflector}

The total deviation $D(b)$, for $b<R$ and in the weak gravity
approximation, is given by eq. (\ref{dout2}) and (\ref{din3}). Setting
$u={b \over R}$, we have
\be
\label{devia3}
D(b)=D_{(in)}(b)+D_{(out)}(b)=2 \ {{\cal G}M \over Rv_{\infty}^2}
\bigl({1-(1-u^2)^{3 \over 2} \over u}\bigr)
\ee

When $v_{\infty} \to c$, this result can be compared to the (GR)
result \cite{Bou_Kan,Adl_Bar}. Notwithstanding  the (ubiquitous)
factor of 2 between classical 
mechanics and (GR), eq. (\ref{devia3}) is in agreement with eq. (7) of
reference \cite{Adl_Bar}. Note that equation (\ref{devia3}) can be
also be written in a way similar to eq. (\ref{devia2}), namely
\be
D(b)=2 \ {{\cal G}M(b) \over bv_{\infty}^2}
\ee
where $M(b)=M (1-\bigl(1-{b^2 \over R^2}\bigr)^{3 \over 2})$ is the
partial deflector mass contained in a cylinder of radius $b$.

For small $b$, one gets $D(b) \sim 3 {b \over f}$. The length
$f={R^2v_{\infty}^2 \over {\cal G}M}$ can be viewed as a focal length; 
its order of magnitude for our model galaxy is $2.5 \ 10^9$
ly, much bigger than $R_G \sim 2 \ 10^4$ ly. Note also that $D(b)$ has a
maximum for $b \simeq 0.93 R_G$, with $D_{max} \simeq 1.6 \ 10^{-5}$ rd.

Using the results of equations (\ref{devia2}),(\ref{devia3}),
we show in Figure 3 the deviation $D(b)$ as a
function of the impact parameter $b$, for opaque and transparent
deflectors. 

\section{Finite distance geometry and multiple trajectories}

\label{geometry}
We now use our results to discuss an experimentally more
relevant situation, where both the source (S) and the observer (E) are
at a finite distance from the deflector ($\Delta$).
The - Euclidean - geometry is shown in Figure 4. We have $r_S=S_0\Delta=SH$,
$r_E=\Delta E$. Given the previous orders of magnitude, the angles
such as $\beta =\widehat{SE\Delta}$, $\theta=\widehat{\Delta EY}$ and
the deviation $D$, are assumed to be small. This implies in particular
that $b=\Delta Y=r_E \tan \theta <<r_E, r_S$. We have from (Euclidean) geometry
\be 
\label{sinus}
{{\rm sin} \widehat{SEY} \over SY}={{\rm sin} \widehat{ESY} \over EY}
\ee
From triangle SHY, one has
\be
SY^2=SH^2+HY^2=r_S^2+\bigl(r_E \tan \theta-(r_E+r_S)\tan \beta\bigr)^2
\ee
yielding for small angles, $SY \simeq r_S (1+O(\theta^2,\beta^2,\theta\beta))$.

From triangle $YE\Delta$ one has
\be
EY^2=r_E^2+\Delta Y^2=r_E^2 (1+\tan^2 \theta)
\ee
yielding for small angles, $EY \simeq r_E (1+O(\theta^2))$.

Plugging these values in eq. (\ref{sinus}), we get to lowest order
in the angles $\theta, \beta,..$
\be
\label{sinus2}
{\widehat{SEY} \over r_S}\simeq {\widehat{ESY} \over
r_E}\simeq {\widehat{SEY}+\widehat{ESY} \over r_S+r_E} \simeq {D \over r_S+r_E}
\ee
Since $\widehat{SEY}=\theta-\beta \simeq {b\over r_E} -\beta$, we
finally obtain 
\be
\label{deviation3}
D(b) \simeq {r_S+r_E \over r_S}({b \over r_E}-\beta)
\ee
Equation (\ref{deviation3}), which expresses the condition that a particle
emitted from (S) reaches the earth, is represented by the dotted lines 
in Figure 3. For an opaque deflector, one may get one or two solutions
for $b$. For a transparent deflector, one may get up to three solutions for
$b$. Rather than studying the full problem as a function of $\beta,
r_S,r_E,...$, we illustrate some particular situations

\subsection{ A generic case}
\label{generic}
This case corresponds to a non zero $\beta$ angle (Figure 4). We focus 
our interest on points (1) and (2) of Figure 3, which are the
intersections of the geometrical equation
\be
\label{deviation4}
D(b) \sim {r_S+r_E \over r_S}({b \over r_E}-\beta) 
\ee
with the $b>R$ gravitational deflection $D(b)$ of Section
\ref{orders} (see Figure 3). Since we have $D(b) \sim
2 \ {{\cal G}M \over bv_{\infty}^2}$, setting $\theta={b \over r_E}$ leads to
\be
\theta^2-\beta\theta -\theta_E^2=0
\ee
where 
\be
\label{Einstein}
\theta_E^2\sim 2 \ {{\cal G}M \over v_{\infty}^2}{r_S \over r_E(r_S+r_E)}
\ee
In this 
case, there are two trajectories in the $S\Delta E$ plane that reach
(E). In photon language, the observer sees two images $(S1)$ and
$(S2)$ of the source (S), on opposite sides of the deflector, with
\be
\label{traj12}
\theta_{1,2}={\beta \pm \sqrt{\beta^2+4\theta_E^2} \over 2}={b_{1,2}
\over r_E}
\ee
For a transparent deflector, one has another image of the source,
corresponding to the point labeled (3) in Figure 3(b).

\subsection{Einstein rings}

This case corresponds to the alignment of (S), ($\Delta$) and (E)
($\beta=0$). Due to the symmetry of revolution around the 
$S\Delta E$ axis, all trajectories on the angular cone
$\theta=\theta_E ={b \over r_E}$ reach the earth. In photon language,
this means that the observer sees a ring image of the 
point source S. For a transparent deflector, one also has a direct
image. For our model galaxy and $r_E \sim r_S \sim
10^9$ ly, a typical value is $\theta_E \sim 10^{-5}$ rd. 

\subsection{The case of a moving deflector}

We briefly consider this case (called microlensing), because of its
experimental relevance. Since a detailed comparison with the
experiments require the use of General
Relativity \cite{Mis_Tho_Whe,Sch_Ehl,Zak_Saz}, we limit our
presentation to orders of magnitude calculations. If the deflector
($\Delta$) moves, with a velocity $v_{\Delta}$, in a direction
perpendicular to the (SE) axis (Figure 4), the above calculations
suggest the following scenario: for $\beta=0$ (ring image), there is a
sudden increase in the signal received by the observer, since two
trajectories only survive for $\beta \ne 0$. Physically the transition 
is gradual, and the observer will receive a gravitationally enhanced
signal when the position of deflector $(\Delta)$ is within a distance
$b_{E} \sim r_E \theta_E$ from the full alignment position of the
previous section \footnote{ A simple way to calculate the
amplification factor is to use the non linear relation $\theta(\beta)$ 
of equation (\ref{traj12}). The flux emitted by an extended source is
proportional to $\beta d\beta$, and the flux received by the observer
is proportional to $\theta d\theta$. The total amplification factor is 
given by ${\cal A}=\vert{\theta d\theta \over \beta
d\beta}\vert_{1}+\vert{\theta d\theta \over \beta d\beta}\vert_{2}$.}
The corresponding time interval is $t_E \sim {b_E \over
v_{\Delta}}$. For distant sources ($r_S >>r_E$), an experimental
situation corresponding to a sun-like deflector, with $v_{\Delta}\sim 
200$ km s$^{-1}$, $r_E \sim 3.6 \ 10^4$ ly and $v_{\infty}=c$, yields an
enhanced signal during an interval $t_E \sim 10^6-10^7$ s, of order 
one month.

\section{Time lags}
\label{time}
\subsection{A simple optical analogy}

We have seen that several trajectories- or light rays- may reach (E)
because of the gravitational deflection. Can one further
extend the mechanical-optical analogy by finding the time lag between
the reception of these trajectories -or light rays- ?

We first consider the transparent deflector for $b<<R$, where we found
$D(b)\sim 3{b \over f}$ in section \ref{orders}. This result may be
compared with the optical deviation of a spherical glass lens, of
radius $\rho$ and of optical index $\nu$ which reads
\be
\label{optik}
D_{opt}=2{\nu-1 \over \nu} \ {b \over \rho}={b \over f_{opt}}
\ee
where $b$ is the impact parameter of the light ray and $f_{opt}={\nu
\rho \over 2(\nu -1)}$ is the focal distance of the lens.
The comparison of the gravitational and optical deviations suggests
that the gravitational deviation may be understood through a
gravitational index $n_{grav}$, with $n_{grav} \ne 1$. 

\subsection{Particule trajectories and geometrical optics}

The preceding remark can be extended and formalized as follows. Energy
conservation for a central potential $U(r)$ reads, in usual polar
coordinates 
\be
E={m \over 2} {\vec v}^2+ mU(r)={m \over 2}({\dot r}^2+r^2 {\dot
\theta}^2)+ mU(r)
\ee
Defining ${\rm tan} \Psi(r)={r(\theta) \over r'(\theta)}$, we have
\be
E={{\vec L}^2 \over 2mr^2{\rm sin}^2\Psi(r)}+mU(r)
\ee
which can be rewritten as
\be
\label{Bouguer}
n_{U}(r) \ r \ {\rm sin}\Psi(r)=({{\vec L}^2 \over 2mE})^{1 \over 2}
\ee
where the ``index'' $n_U(r)$, associated to the potential $U(r)$ is
given by $n_U(r)=\sqrt{1-{mU(r) \over E}}$. Equation (\ref{Bouguer})
is analogous to Bouguer's relation for the propagation of light rays
in a spherically symmetric medium of index $n_U(r)$ \cite{Born}.

\bigskip
We will illustrate this analogy with the case $b>R$, where
$U(r)=-{{\cal G}M \over r}$. The associated gravitational index reads
\be
\label{index}
n_{grav}(r)=\sqrt{1+{2{\cal G}M \over v_{\infty}^2r}} \simeq {1+{{\cal
G}M \over v_{\infty}^2r}}  
\ee
where we have used the weak gravity approximation.
Defining an analog $L_{grav}$ of the optical path, we may express the
time lag $\delta t_{12}$ between the reception on earth of trajectories
(1) and (2) of section \ref{generic} as

\be
\label{fermat1}
\delta t_{12}={L_2-L_1 \over v_{\infty}}={\int_{(2)}
n_{grav}(r) \ ds_2 -\int_{(1)} n_{grav}(r) \ ds_1
\over v_{\infty}} 
\ee
From eqs. (\ref{index}) and (\ref{fermat1}),  one finds that $\delta
t_{12}$ is the sum of a geometrical part $\delta t_{geom}={\int_{(2)}
\ ds_2-\int_{(1)} \ ds_1 \over v_{\infty}}$ and of a gravitational
part 
\be
\label{grav2}
\delta t_{grav}={{\cal  G}M \over v_{\infty}^3}\bigl(\int_{(2)}
{ds_2 \over r}-\int_{(1)} {ds_1 \over r}\bigr)
\ee
The full calculation of the integrals in (\ref{grav2}) (see eq. 8.30, p.240 of
\cite{Sch_Ehl}), yields a result that depends only logarithmically on
the geometrical parameters ($\beta, r_S,...$). We therefore estimate $
\delta t_{grav} \sim {{\cal G}M \over v_{\infty}^3}$, up to a numerical 
factor of order one.
Setting $v_{\infty}=c$, we find
$\delta t_{grav} \sim 10^{-5}$ s for the sun, and $\delta
t_{grav} \sim 10^{7}$ s for our model galaxy. Experiments that
confirm the double reception of the same ``signal'', with a
gravitational time lag of order several months, can be found in
references \cite{Mis_Tho_Whe,Sch_Ehl,Zak_Saz,Adl_Bar,Wamb}.

Finally, it of interest to note that a Fermat approach
to the weak gravity approximation of (GR) yields an equation similar to
equation (\ref{index}), with a (GR) index $n_{GR} \sim 1+2 \ {{\cal
G}M \over c^2r}$ \cite{Nan_Hel,Als}.

\section{Conclusion}
We have studied, at a qualitative level, a classical mechanical
introduction to gravitational lens phenomena. This approach rests
on the fact that the gravitational deflection of a massive particle by
a deflector is independent of the particle mass. 
It is only an approximation to the theory of General Relativity
\cite{Nan_Hel,Als}, 
but I believe that this ``$\vec F=m \vec a$ optics ''
\cite{Eva_Ros,Bel_Rod} brings together in a very pedagogical way problems of 
different origins. In particular, we have derived orders of magnitude
for the weak gravity case,
that can be compared -up to a factor 2- to the correct (GR)
results. As a caveat, we have nevertheless pointed out that the
particle velocity may become (weakly) supraluminal, and this (weak)
inconsistency with relativity is to be kept in mind.

 Beside the study of General Relativity, the interested
student can carry further the present approach in several ways. I will
only quote here the modeling of gravitational lenses by optical lenses
of the appropriate shape \cite{Adl_Bar}, or the link between equations 
(\ref{newton1}) and (\ref{newton2}) stemming from conformal
transformations \cite{Mitt_Ste}. 

\vskip 2mm

It is a pleasure to thank F. Bernardeau for discussions.

\newpage

\centerline{\bf Figure Captions}
\vskip 10mm
{\bf Figure 1:} The deflection geometry for $b>R$, with
(S) and (E) at infinity. The vectors $\vec u_i$ and $\vec u_f$ are the 
unit vectors of the initial (emission)  and final (reception)
directions. The deviation $D=\arccos (\vec u_i \cdot \vec u_f) $ is a
function of the impact parameter $b$. The minimum distance approach is
$r_0$, and $r_0 \simeq b$ for small $D$.

\vskip 10mm

{\bf Figure 2:} The deflection geometry for the transparent deflector

(a) phase (out) for $b<R$. The vector $\vec u_{in}$ is the unit vector 
along the velocity as the particule enters the deflector at point M
($\vec v_{in}=v_0 \vec u_{in}$). The corresponding deviation is given by
$D_{out}^{(1)}=\arccos (\vec u_i \cdot \vec u_{in})$

(b) phase (in) for $b<R$.  The particle enters deflector ($\Delta$) at
point M, and exits at point N. The OX and OY axes are the eigenvectors
of the Runge-Lenz tensor ${\bf A}$. Points M and N are symmetric w.r.t. OY.
For clarity purposes, the direction of $\vec u_{in}$ has been rotated
with respect to Figure 2(a).

\vskip 10mm
{\bf Figure 3:} The qualitative variation of $D(b)$ (a) opaque deflector
(b) transparent deflector (note the maximum for $b \sim R_G$). The
dotted lines represent various cases of  equation
(\ref{deviation3}). Solutions (1) and (2) correspond to trajectories
(1) and (2) of Figure 4.

\vskip 10mm

{\bf Figure 4:} A typical geometry for finite distance gravitational
deflection. Trajectories (1) and (2), deflected by $(\Delta)$, reach
the earth (E): the source (S) has two images (S1) and (S2). For 
a transparent deflector, one has a third image of the source
( see point (3) in Figure 3(b)), corresponding to a trajectory
entering ($\Delta$).

\newpage

\begin{figure}
\begin{center}
\includegraphics[height=6cm]{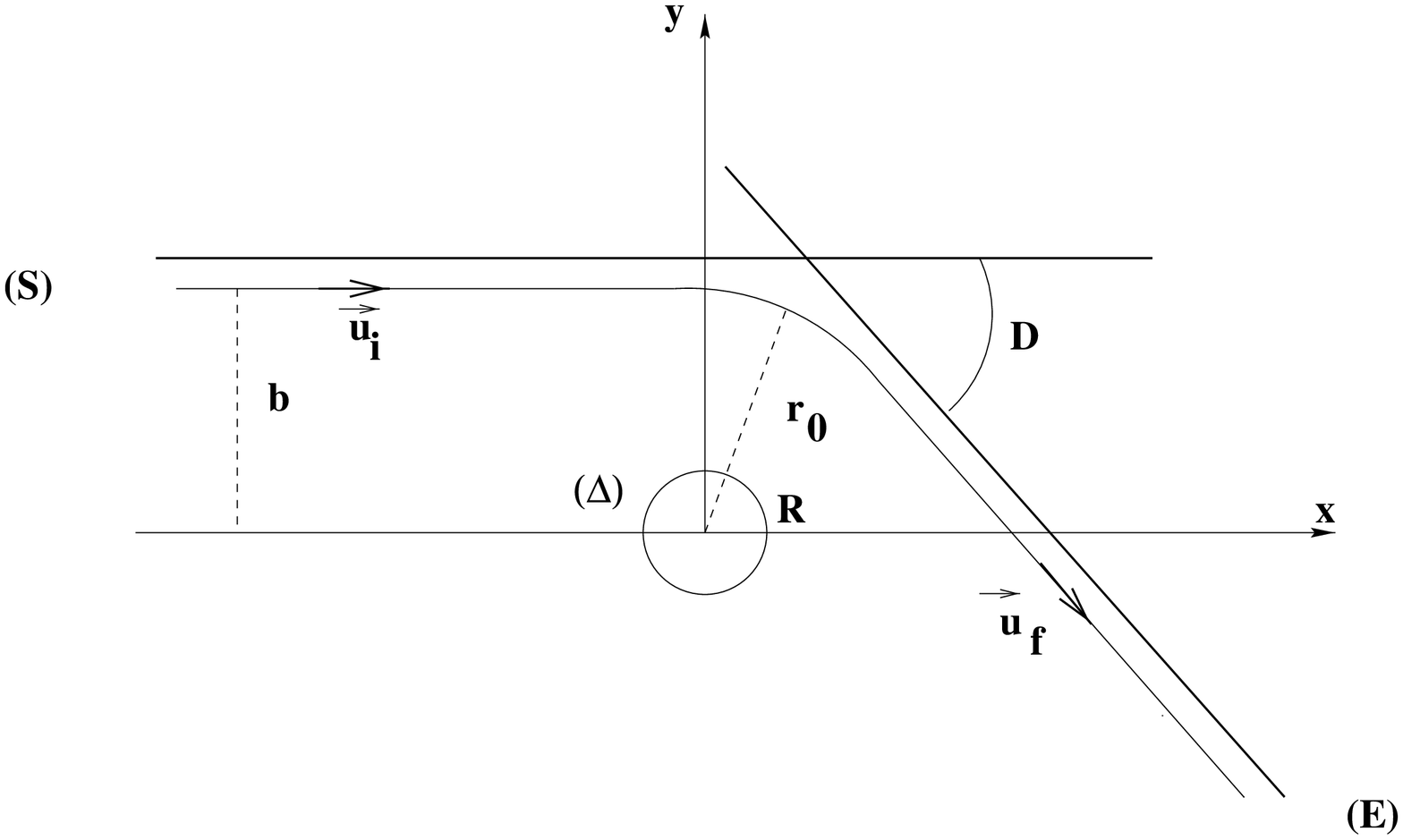}
\end{center}
\centerline{Figure 1}
\vskip 20mm

\begin{center}
\includegraphics[height=6cm]{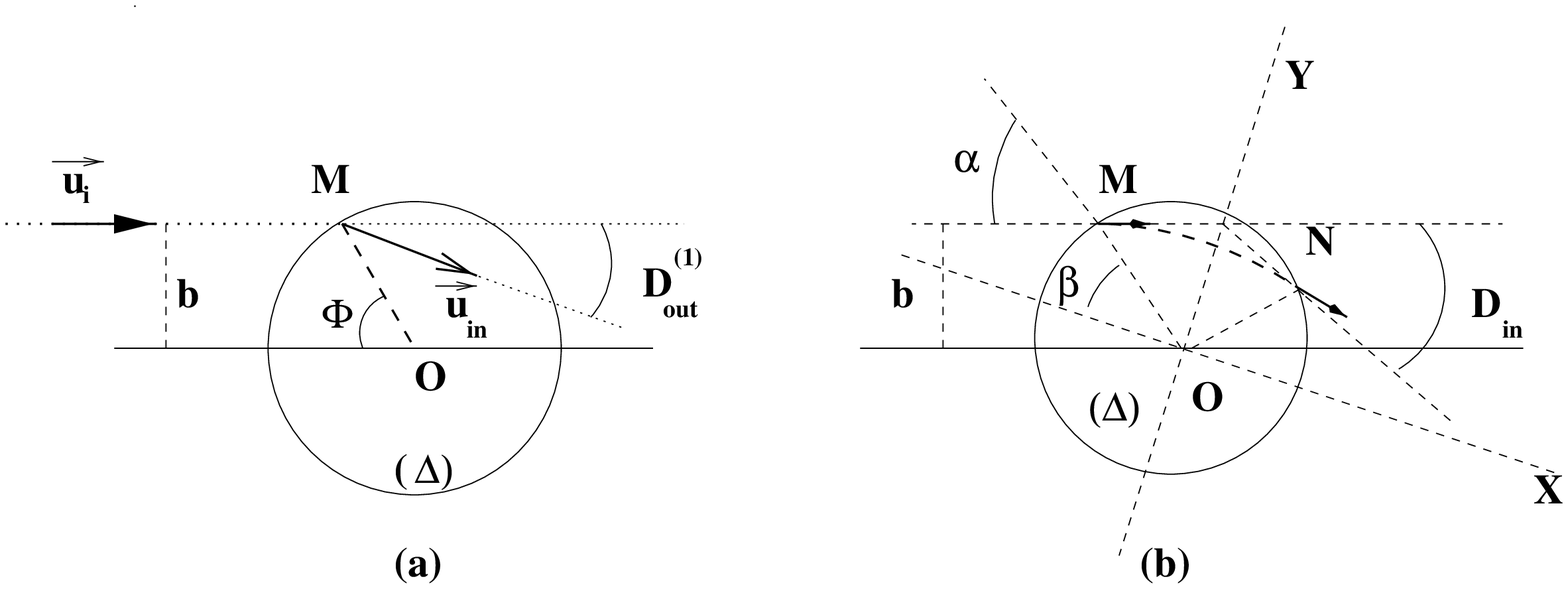}
\end{center}
\centerline{Figure 2}
\end{figure}
\vskip 10mm

\newpage

\begin{figure}
\begin{center}
\includegraphics[height=8cm]{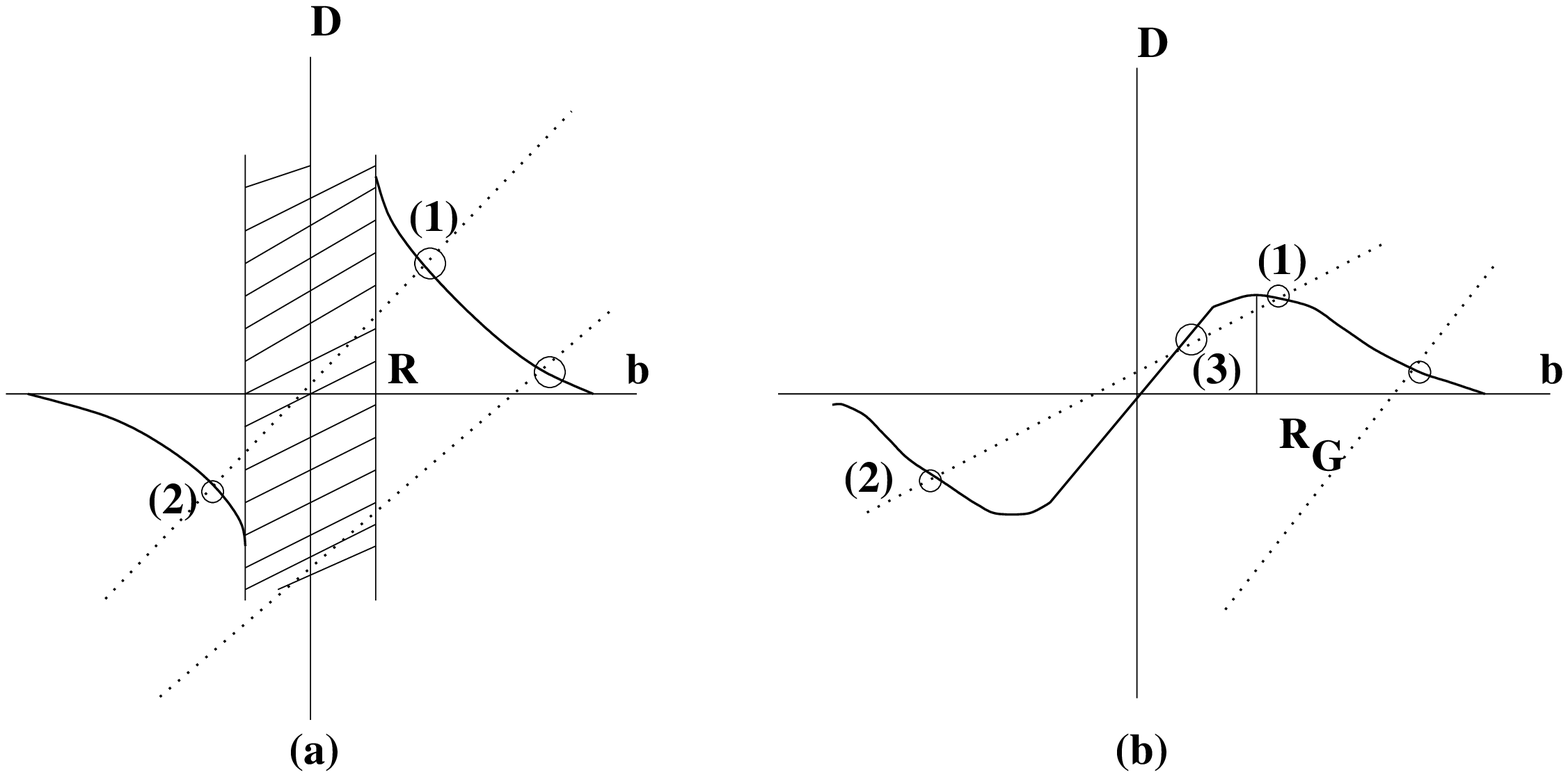}
\end{center}
\centerline{Figure 3}
\end{figure}
\vskip 150mm

\newpage

\begin{figure}
\begin{center}
\includegraphics[width=13cm]{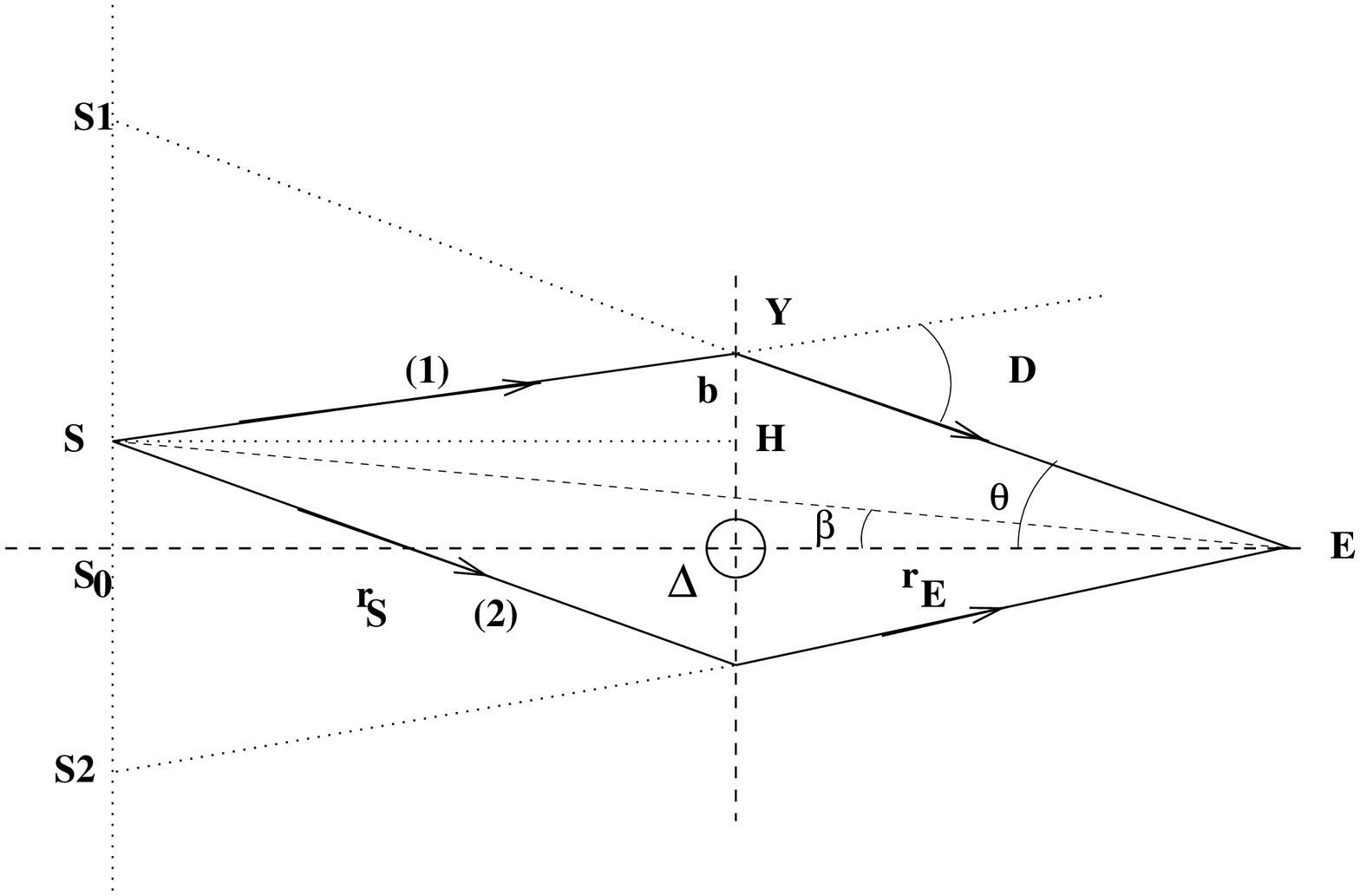}
\end{center}
\vskip 10mm
\centerline{Figure 4}
\end{figure}

\end{document}